\documentclass[journal]{IEEEtran}
\ifCLASSINFOpdf
\else
\fi
\usepackage[cmex10]{amsmath}
\usepackage{amsfonts}
\usepackage{amsthm}
\usepackage{amssymb}
\usepackage{latexsym}
\usepackage[latin1]{inputenc}

\begin{document}

\newtheorem{theo}{Theorem}[section]
\newtheorem{lemma}[theo]{Lemma}
\newtheorem{cor}[theo]{Corollary}

\newtheorem{definition}[theo]{Definition}
\newtheorem{example}[theo]{Example}
\newtheorem{remark}[theo]{Remark}
\newtheorem{conj}[theo]{Conjecture}
\newtheorem{prop}[theo]{Proposition}
\newtheorem{reference}{}
\newcommand{\wh}{\widehat}
\newcommand{\ol}{\overline}

\newcommand{\Hom}{{\rm Hom}\,}           
\newcommand{\Aut}{{\rm Aut}}             
\newcommand{\zzz}{\mathbb{Z}}        
\newcommand{\car}{{\rm char}}
\def\F{{\mathbb F}}

%
\title{$G$-equivalence in group algebras and minimal abelian codes}
%
%
%

\author{Raul~Antonio~Ferraz,
        Marinês Guerreiro,
        and~César~Polcino~Milies,
\thanks{R.A. Ferraz and C. Polcino Milies are with the Instituto de Matemática e Estatística, Universidade de São Paulo, Caixa Postal 66281, CEP 05314-970 - São Paulo - SP (Brasil), supported by CNPq, Proc. 300243/79-0(RN) and FAPESP, Proc. 09/52665-0. E-mail: raul@ime.usp.br and polcino@ime.usp.br.}
\thanks{M. Guerreiro is with Departamento de Matemática, Universidade Federal de Vi\c cosa, CEP 36570-000 - Vi\c cosa-MG (Brasil), supported by 
CAPES, PROCAD 915/2010 and FAPEMIG, APQ CEX 00438-08. E-mail: marines@ufv.br.}
\thanks{This work was presented in part at the  2011 IEEE Information Theory Workshop (ITW), Paraty-RJ (Brasil) 16-20 October,  2011. }
\thanks{Manuscript received ???, 20012; revised ???, 2012.}}

%
%

\markboth{IEEE Transactions of Information Theory,~Vol.~??, No.~??, January~20??}%
{Shell \MakeLowercase{\textit{et al.}}: Bare Demo of IEEEtran.cls for Journals}
%



\maketitle

\begin{abstract}
Let $G$ be a finite abelian group and $\F$ a field such that $\car(\F)\not | \;|G|$. Denote by
$\F G$ the group algebra of $G$ over $\F$.
A (semisimple) abelian code is an ideal of $\F G$.  
Two codes ${\cal I}_1$ and ${\cal I}_2$  of $\F G$ are 
{\em $G$-equivalent} if there exists an automorphism $\psi$ of $G$ whose linear extension 
to $\F G$ maps ${\cal I}_1$ onto ${\cal I}_2$.

In this paper we give a necessary and sufficient condition for minimal abelian codes to be $G$-equivalent and
 show how to correct some results in the literature.
\end{abstract}


%
\IEEEpeerreviewmaketitle

\section{Introduction}
%
%
%
%
\IEEEPARstart{L}{et} $G$ be a finite group and $\F$ a finite field such that $\car(\F)\not\!\! | \;|G|$. 
Two ideals ${\cal I}_1$ and ${\cal I}_2$  of the group algebra $\F G$ are said to be
{\bf\em $G$-equivalent} if there exists an automorphism $\psi$ of $G$ whose linear extension 
to $\F G$ maps ${\cal I}_1$ onto ${\cal I}_2$.

This definition was introduced R.L. Miller~\cite{miller} in the context of Coding Theory. 
S.D. Berman~\cite{berman} and, independently, F.J. MacWilliams~\cite{mw} defined abelian codes 
as ideals in finite abelian group algebras and R.L. Miller~\cite{miller} used $G$-equivalence to compare codes with the
same weight distribution. 

In this paper, we address the problem of determining $G$-equivalence of minimal ideals in semisimple 
abelian group algebras and prove that the $G$-equivalence classes of minimal ideals depend on the structure of the lattice
of the subgroups of $G$.

In Section~\ref{section2} we prove preliminary results about idempotents and in
Section~\ref{sec3} we establish a correspondence between $G$-equivalence classes of minimal abelian ideals in $\F G$
 and certain isomorphism classes of subgroups of $G$.
In Section~\ref{positive}, we use these facts to show that some of the results of~\cite{miller} are not correct and, in the final section,
 we exhibit particular cases for which such results hold.

 

\section{Subgroups and Idempotents}\label{section2}

The irreducible central idempotents of the rational group algebra $\mathbb{Q}G$ were 
computed in~\cite[Theorem~1.4]{goodcesar} in the case when $G$ is abelian; in \cite[Theorem 2.1]{jlp} 
when $G$ is nilpotent; in~\cite[Theorem 4.4]{OdRS} when $G$ is abelian-by-supersolvable and 
in~\cite[Theorem 7]{BdR} an algorithm to write the primitive idempotents in given. 

In what follows, we shall establish a correspondence between primitive idempotents of $\F G$ and certain subgroups of $G$.

\begin{definition}
Let $G$ be a group. A subgroup $H$  of $G$ is said a \textbf{co-cyclic subgroup} if the factor group $G/H\neq 1$ is  cyclic.
\end{definition}
We use the notation
$$
{\mathcal{S}}_{cc}(G)=\{H\, |\, H \mbox{ is a co-cyclic subgroup of } G\}.
$$

We recall the following results that are used throughout this paper.

Let $G$ be a finite abelian $p$-group and $\F$  a field such that  $\car(\F)\not\!\!\! |\;|G|$.
Given a subgroup $H$ of $G$, denote $\wh H = \frac{1}{|H|}\sum_{h\in H}h$ and, for an element $x\in G$, set $\wh x = \wh{\left< x\right>}$.
For each co-cyclic subgroup $H$ of $G$, we can construct an idempotent of $\F G$.
In fact, we remark that, since $G/H$ is a cyclic $p$-group, there exists a unique subgroup $H^{\sharp}$ of $G$ containing $H$ such that $|H^{\sharp}/H|=p$.
Then $e_H=\wh H - \wh{H^{\sharp}}$ is an idempotent and we consider the set 
\begin{equation}\label{idempabelian}
\{\wh G\}\cup \{ e_H=\wh H - \wh{H^{\sharp}}\,|\,H\in {\mathcal{S}}_{cc}(G), G/H \neq \{ 1\}\}.
\end{equation}

In the case of a rational abelian group algebra $\mathbb{Q} G$, the set above is the set of primitive central idempotents~\cite[Theorem~1.4]{goodcesar}. Also, the following results holds.

\begin{theo}{\em \cite[Lemma~5]{FM}} \label{3.44}
Let $p$ be a prime rational integer and $G$ a finite abelian group of exponent $p^n$ and $\F_q$ a finite field such that $p\!\not | \,q$. Then~\eqref{idempabelian} 
  is a set of pairwise orthogonal idempotents of $\F_q G$ whose sum is equal to $1$.
\end{theo}

\begin{theo}{\em \cite[Theorem~4.1]{FM}} \label{3.4} Under the hypotheses above, 
the set~\eqref{idempabelian} is the set of primitive idempotents of $\F_q G$ if and only  if    $o(\bar{q}) = \phi (p^n)$ in $U(\zzz_{p^n})$,
  where $\varphi$ denotes Euler´s totient function.
\end{theo}

We shall repeatedly use the following rather obvious fact.

\begin{lemma}\label{unicosbg}
Let $G$ be a finite abelian $p$-group and $H\leq G$. Then $G/H$ is a cyclic group if and only if there exists a unique subgroup  $L$ such that $H< L \leq G$ and $[L:H]=p$.
\end{lemma}

In the sequel, for a finite abelian group $G$, we write $G=G_{p_1}\times \cdots \times G_{p_t}$, where
$G_{p_i}$ denotes the $p_i$-Sylow subgroup of $G$, for the distinct positive prime numbers $p_1,\ldots, p_t$.

\begin{lemma}\label{co-cyclic}
Let $G=G_{p_1}\times  \cdots \times G_{p_t}$ be a finite abelian group and $H\in {\mathcal{S}}_{cc}(G)$.
Write $H=H_{p_1}\times  \cdots \times H_{p_t}$, where $H_{p_i}$ is the $p_i$-Sylow subgroup of $H$. Then 
each subgroup $H_{p_i}$  is co-cyclic in $G_{p_i}$, $1\leq i\leq t$.
\end{lemma}
\begin{IEEEproof}
For $H\in {\mathcal{S}}_{cc}(G)$, the quotient $G/H\cong G_{p_1}/H_{p_1}\times \cdots \times G_{p_t}/H_{p_t}$
is cyclic, hence each factor $G_{p_i}/H_{p_i}$ must be cyclic. Therefore, $H_{p_i}\in {\mathcal{S}}_{cc}(G_{p_i})$,  $1\leq i\leq t$.
\end{IEEEproof}

With the notation above, for each $H\in {\mathcal{S}}_{cc}(G)$, define an idempotent $e_H\in \F G$ as follows. For each $1\leq i\leq t$,
either $H_{p_i}=G_{p_i}$ or there exists a unique subgroup $H_{p_i}^{\sharp}$ such that $[H_{p_i}^{\sharp}:H_{p_i}]=p_i$. Thus, let
$e_{H_{p_i}}=\wh{G_{p_i}}$ or $e_{H_{p_i}}=\wh{H_{p_i}}-\wh{H_{p_i}^{\sharp}}$, respectively, and define
\begin{equation}\label{idempH}
e_H = e_{H_{p_1}}e_{H_{p_2}}\cdots e_{H_{p_t}}.
\end{equation}

For any other $K\in {\mathcal{S}}_{cc}(G)$, with $K\neq H$, we have $K_{p_i}\neq H_{p_i}$, for some $1\leq i\leq t$, hence $e_{H_{p_i}}e_{K_{p_i}}=0$ and so $e_He_K=0$. Thus, we have the following.

\begin{prop}\label{baseB}
Let $G$ be a finite abelian group and $\F$ a field such that $\car(\F)\not |\;|G|$. 
Then $${\mathcal{B}}=\{ e_H\,|\,H \in {\mathcal{S}}_{cc}(G) \}$$
is a set of orthogonal idempotents of $\F G$.
\end{prop}

A similar construction of idempotents for rational group algebras of abelian groups is given in~\cite[Section VII.1]{goodcesar}. For the rational case, these idempotents are primitive while for finite fields this is usually not true.

To study the $G$-equivalence of ideals, we need to understand how the group of automorphisms $\Aut(G)$ acts on the
lattice of the subgroups of $G$ and hence on the idempotents in the group algebra which arise from these subgroups.
From now on, we use the same notation for an automorphism of the group $G$ and its linear extension to
the group algebra $\F G$.

\begin{lemma}\label{invaut}
Let $G$ be a finite abelian group, $H\in {\mathcal{S}}_{cc}(G)$ and $e_H$ its corresponding idempotent defined as in~\eqref{idempH}. Then, for any $\psi\in \Aut(G)$,  we have $\psi(e_H)=e_{\psi(H)}$.
 \end{lemma}
\begin{IEEEproof}
By Lemma~\ref{co-cyclic}, $H=H_{p_1}\times H_{p_2}\times \cdots \times H_{p_t}$, where $H_{p_i}$ is the $p_i$-Sylow subgroup of $H$ which is
co-cyclic in $G_{p_i}$ (the $p_i$-Sylow subgroup of $G$), for each $1\leq i\leq t$. Since $\psi\in\Aut(G)$, $\psi(H)=\psi(H_{p_1})\times\psi(H_{p_2})\times \cdots \times \psi(H_{p_t})$. Then each $\psi(H_{p_i})$ is the $p_i$-Sylow subgroup of $\psi(H)$ and is also co-cyclic in $G_{p_i}$. Hence $\psi(H_{p_i}^{\sharp})=\psi(H)_{p_i}^{\sharp}$ and the result follows.
\end{IEEEproof}

\begin{lemma}\label{sum=1}
Let $G$ be a finite abelian group and $\F$ a field such that $\car(\F)\not |\;|G|$. 
Then, in the group algebra $\F G$, we have:
\begin{equation}\label{soma=1}
1= \wh G\,+\,\sum_{H\in {\mathcal{S}}_{cc}(G)} e_H.
\end{equation}
\end{lemma}
\begin{IEEEproof}
Let $G=G_{p_1}\times G_{p_2}\times \cdots \times G_{p_t}$, with
$G_{p_i}$ the $p_i$-Sylow subgroup of $G$, $1\leq i\leq t$. By Theorem~\ref{3.44}, we have
\begin{equation}\label{somaHi}
1 = \wh G_{p_i}\,+\,\sum_{H_{p_i}\in {\mathcal{S}}_{cc}(G_{p_i})} (\wh{H_{p_i}}-\wh{H_{p_i}^{\sharp}}).
\end{equation}
Thus:
\begin{eqnarray*}\label{productHi}
1 & = & \prod_{i=1}^{t}\left(\wh G_{p_i}\,+\,\sum_{H_{p_i}\in {\mathcal{S}}_{cc}(G_{p_i})} (\wh{H_{p_i}}-\wh{H_{p_i}^{\sharp}})\right) \\
& = & \sum e_{H_{p_1}}e_{H_{p_2}}\cdots e_{H_{p_t}},
\end{eqnarray*}
where either $e_{H_{p_i}}=\wh{G_{p_i}}$ or $e_{H_{p_i}}=\wh{H_{p_i}}-\wh{H_{p_i}^{\sharp}}$, $1\leq i\leq t$.
Therefore, $$1= \wh G\,+\,\sum_{H\in {\mathcal{S}}_{cc}(G)} e_H.$$
\end{IEEEproof}

\begin{lemma}\label{uniqueco-cyc}
Let $G$ be a finite abelian group and $\F$ a field such that $\car(\F) \not\!  |\;|G|$. For each primitive idempotent $e\in\F G$, there
exists a unique $H\in {\mathcal{S}}_{cc}(G)$ such that $e\cdot e_H = e$ and $e\cdot e_K=0$, for any other $K\in {\mathcal{S}}_{cc}(G)$.
\end{lemma}
\begin{IEEEproof}
By Lemma~\ref{sum=1}, $1 = \wh G\,+\,\displaystyle\sum_{H\in {\mathcal{S}}_{cc}(G)} e_H$. Multiplying by $e$, we have:
\begin{equation}\label{last}
e = e\left(\wh G\,+\,\sum_{H\in {\mathcal{S}}_{cc}(G)} e_H\right)=e\cdot\wh G\,+\,\sum_{H\in {\mathcal{S}}_{cc}(G)} e\cdot e_H.
\end{equation}
As $e_H\cdot e_K=0$, for $H\neq K$ co-cyclic subgroups of $G$, the right hand side of~\eqref{last} is a sum of orthogonal idempotents. Therefore,
as $e$ is a primitive idempotent, only one summand is non-zero and this proves the lemma.
\end{IEEEproof}

Set:
$$ \mathcal{P}(\F G)=\{e\in \F G \,|\, e \mbox{ is a primitive idempotent in } \F G\}.$$
Under the same hypotheses of Lemma~\ref{uniqueco-cyc}, the following map is well-defined:
\begin{equation}\label{map}
\begin{array}{clc}
  \Phi\; : \;\mathcal{P}(\F G) & \longrightarrow & {\mathcal{S}}_{cc}(G) \\
 e &\longmapsto & \Phi(e)=H_e,
\end{array}
\end{equation}
where $H_e$ is the unique co-cyclic subgroup of $G$ such that $e\cdot e_{H_e}=e$.

\begin{theo}\label{somaeH}
Let $G$ be a finite abelian group, $\F$ a field such that $\car(\F) \not |\;|G|$ and $H\in {\mathcal{S}}_{cc}(G)$.
Then $e_H$ is the sum of all primitive idempotents $e\in \mathcal{P}(\F G) $ such that $\Phi(e)=H$.
\end{theo}
\begin{IEEEproof}
Write $1=\displaystyle\sum_{e\in\mathcal{P}(\F G)} e$. Then
$$ 
e_H = \sum_{e\in\mathcal{P}(\F G)} e_H e = \sum_{\Phi(e)\neq H} e_H e\,+\,\sum_{\Phi(e)=H} e_H e
= \sum_{\Phi(e)=H} e.
$$
\end{IEEEproof}
\begin{remark}\label{rem1}
For a primitive idempotent $e\in \F G$ and its correspondent subgroup $H=H_e$, as in~\eqref{map}, we have:
$$
e\wh K = e\;\Leftrightarrow \; K\leq H\;\mbox{ and }\; e\wh K = 0\; \Leftrightarrow \;K\not\leq H.
$$
\end{remark}
\begin{IEEEproof}
By~\eqref{idempH}, $e_H = e_{H_{1}}e_{H_{2}}\cdots e_{H_{t}}$, where either $e_{H_{i}}=\wh{G_{i}}$ or $e_{H_{i}}=\wh{H_{i}}-\wh{H_{i}^{\sharp}}$,
where $H_{i}$ denotes the $p_i$-Sylow of $H$, $1\leq i\leq t$. 

Let $K\leq H$. Then $K_i\leq H_i\leq H_i^{\sharp}$, where $K_{i}$ is the $p_i$-Sylow of $K$. Hence $\wh{K_i}\wh{H_i} = \wh{H_i}$ and $\wh{K_i}\wh{H_i^{\sharp}} = \wh{H_i^{\sharp}}$.
Thus either $\wh{K_i}\wh{G_{i}}=\wh{G_{i}}$ or $\wh{K_i}e_{H_i}=\wh{K_i}(\wh{H_i}-\wh{H_i^{\sharp}})=\wh{H_i}-\wh{H_i^{\sharp}}=e_{H_i}$. Therefore,
$$e\wh K=e e_H\wh K = e(e_{H_{1}}e_{H_{2}}\cdots e_{H_{t}})(\wh{K_1}\wh{K_2}\cdots \wh{K_t})=e e_H=e.
$$
If $K\not\leq H$, then there exists at least one $K_i\not\leq H_i$. In this case, $H_i^{\sharp}\subset K_iH_i$, hence  $\wh{K_i}\wh{H_i^{\sharp}} = \wh{K_i}\wh{H_i}$
and $\wh{K_i}e_{H_i}=\wh{K_i}(\wh{H_i}-\wh{H_i^{\sharp}})=\wh{K_i}\wh{H_i}-\wh{K_i}\wh{H_i}=0$. This proves the remark.
\end{IEEEproof}

\begin{prop}\label{quotient}
Let $G$ be a finite abelian group and $\F$ a field such that $\car(\F) \not\! |\;|G|$. If $e\in\F G$ is a primitive idempotent,
then $G e\cong G/H_e$.
\end{prop}
\begin{IEEEproof}
It is clear that $G e$ is a group. Consider $\pi : G\longrightarrow G e$ given by $\pi(g)=ge$. 
Clearly
$H_e\subset ker(\pi)$. For $g\in ker(\pi)$, $\pi(g)=e$ implies $ge=e$, hence $\wh{\langle g\rangle}e=e$ and, by Remark~\ref{rem1},
$\langle g\rangle\subset H_e$. Therefore, $g\in H_e$ and $ker(\pi)=H_e$, proving the lemma.
\end{IEEEproof}

Let $G$ be a finite abelian group. We recall that its {\bf character group} 
is $G^*=\Hom (G,\mathbb{C}^*)$, with multiplication defined by
$(f\cdot g) (x)=f(x)\cdot g(x)$, for all $f,\, g\in G^*$ and $x\in G$.
Also for a subgroup $H$ of $G$, define
$$H^{\perp}= \{ f\in G^* | f(h)=1, \mbox{ for all } h\in H\}.$$

The following facts on character groups 
can be found in~\cite[Chapter 10]{rotman} and will be
used in the next sections.

\begin{theo}\label{annihil}
Let $G$ be a finite abelian group. Then:

(1) $G^* \cong G$.

(2) If $H \leq G$, then $G$ contains a subgroup isomorphic to $G/H$.

(3) $H^{\perp}$ is a subgroup of $G^*$ and $H^{\perp}\cong (G/H)^*$.

(4) Let $ {\cal S}(G)$ be the set of all subgroups of a group $G$.
Then the map
$$
\begin{array}{lclll}\label{psiHo}
\Psi & : & {\cal S}(G) & \longrightarrow & {\cal S}(G^*) \nonumber\\
     &   & H & \longmapsto & H^{\perp}
\end{array}
$$
is a bijection and satisfies:

(a) If $H\subset K$ then $ H^{\perp}\supset K^{\perp}$ (with proper inclusion preserved).

(b) If $H$ is a cyclic subgroup of order $p^s$ in a finite abelian $p$-group $G$ of exponent $p^r$, with $r\geq s$, then
$H^{\perp}$ is a co-cyclic subgroup of $G$ such that $G^*/H^{\perp}\cong C_{p^s}\cong H$.

(c) If $H$ is a co-cyclic subgroup such that
$G/H\cong C_{p^s}$ in a finite abelian \linebreak $p$-group $G$ of exponent $p^r$, with $r\geq s$, then $H^{\perp}$ is a cyclic subgroup isomorphic to $C_{p^s}$.
\end{theo}

\section{$G$-isomorphisms of subgroups}\label{sec3}

We say that two subgroups $H$ and $K$ of a group $G$ are {\bf\em $G$-isomorphic} if there exists an automorphism $\psi\in \Aut(G)$
such that $\psi(H)=K$.

Notice that isomorphic subgroups are not necessarily $G$-isomorphic. For example, for a prime $p$,
if $G=\langle a\rangle\times \langle b\rangle$ with $o(a)=p^2$ and $o(b)=p$, then $\langle a^p\rangle$ and $\langle b \rangle$ are isomorphic but not $G$-isomorphic, since $\langle b\rangle$ is contained properly only in $\langle a^p\rangle\times \langle b\rangle$ and
$\langle a^p\rangle$ is contained in $\langle a\rangle$ and in $\langle a^ib\rangle$, for all $1\leq i\leq p-1$.

For finite abelian groups, Propositions~\ref{prop1},~\ref{prop2} and~\ref{recprop1} below establish a correspondence between
$G$-equivalent minimal ideals and $G$-isomorphic subgroups.

\begin{prop}\label{prop1}
Let $G$ be a finite abelian group and $\F$ a field such that $\car(\F)\not\! |\;|G|$.
If $e,\,e_1\in \mathcal{P}(\F G)$ are such that $\psi(e)=e_1$, for some automorphism $\psi\in \Aut(G)$ linearly extended to $\F G$, then
$$ \psi(H_e)=H_{\psi(e)}=H_{e_1},$$
i.e., $H_e$ and $H_{e_1}$ are $G$-isomorphic.
\end{prop}
\begin{IEEEproof}
Given $e,\,e_1\in \mathcal{P}(\F G)$, from Lemma~\ref{uniqueco-cyc}, there exist $H_e, \, H_{e_1} \in {\mathcal{S}}_{cc}(G)$ such that
$$ e\cdot e_{H_e} = e \qquad \mbox{ and } \qquad e_1\cdot e_{H_{e_1}} = e_{1}.$$
Hence, by Lemma~\ref{invaut},
\begin{equation}\label{ig}
 \psi(e) = \psi (e\cdot e_{H_e}) = \psi(e)\psi(e_{H_e})= \psi(e)\cdot e_{\psi(H_e)}.
\end{equation}
Since $\psi(e)$ is also a primitive idempotent in $\F G$, Lemma~\ref{uniqueco-cyc} shows that there exists a unique
subgroup in $G$ satisfying~\eqref{ig}, hence
$\psi (H_e)=H_{\psi(e)}$. As $\psi(e)=e_1$, we have $\psi(H_e)=H_{e_1}$.
\end{IEEEproof}

The converse of the Proposition~\ref{prop1} is also true and it will be proved in Proposition~\ref{recprop1} as we still need more information. We set
$$
{\mathcal{L}}\Aut(G) = \{ \psi\in\Aut(G)\,|\,\psi(H)=H, \mbox{ for all } H\leq G \}.
$$

\begin{lemma}\label{gemgr}
Let $G$ be a finite abelian group, $g\in G$ and $r\in \mathbb{N}$ with $\gcd (r,\,o(g))=1$. Then there exists
$\psi\in {\mathcal{L}}\Aut(G)$ such that $\psi(g)=g^r$.
\end{lemma}
\begin{IEEEproof}
We can write $|G|=k\cdot s$ such that $k$ and $o(g)$ have the same prime divisors and
 $\gcd ( s,\,o(g))=\gcd ( k, \,s)=1$. By the Chinese Remainder Theorem, there exists
$x\in\mathbb{Z}$ satisfying
$$ x\equiv r\,(\mod k)\quad \mbox{ and } \quad x\equiv 1\, (\mod s).$$
This implies $\gcd ( x,\, k) = \gcd ( x, \,s)=1$, hence $\gcd (x, \,|G|)=1$.
Therefore, defining $\psi: G\longrightarrow G$ by $\psi(h)=h^x$, we have $\psi\in {\mathcal{L}}\Aut(G)$  and  $\psi(g)=g^x=g^r$.
\end{IEEEproof}

\begin{lemma}\label{locautchar}
Let $G$ be a finite abelian group and $\psi\in\Aut(G)$.
Then $\psi\in{\mathcal{L}}\Aut(G)$ if and only if
there exists $r\in\mathbb{N}$ such that $\gcd (r,|G|)=1$ and
$\psi(g)=g^r$, for all $g\in G$.
\end{lemma}
\begin{IEEEproof}
Suppose $\psi\in{\mathcal{L}}\Aut(G)$. For each $g\in G$, we have $\psi(\langle g\rangle)= \langle g\rangle$, hence
$\psi(g)=g^{r_g}$, for some $r_g\in\mathbb{N}$ such that $\gcd ( r_g,|G|)=1$.

Write $G\cong C_{s_1}\times C_{s_2}\times \cdots \times C_{s_t}$, with
$C_{s_j}=\langle a_j\rangle$ of order $s_j | |G|$ and $s_{j+1}|s_j$,  $1\leq j\leq t$.
Thus $\psi(a_j) = a_j^{r_j}$, with $\gcd ( r_j,|G|)=1$,  $1\leq j\leq t$.
As $s_{j} | s_1$,  then $r_j\equiv r_1 \,(\mod s_j)$,  $1\leq j\leq t$, and  $r_g\equiv r_1\, (\mod s_j)$, 
 $1\leq j\leq t$.
Therefore, we can take $r_g=r_1$. This proves the lemma, since the converse is clear.
\end{IEEEproof}

\begin{lemma}\label{basepA}
Let $G$ be a finite abelian group and $\F$ a field such that $\car(\F)\not\! |\,\;|G|$.
Then
${\mathcal{B}}=\{ e_H\,|\,H \in {\mathcal{S}}_{cc}(G) \}$ is both a basis for the algebra
$$
{\mathcal{A}}=\{ \alpha \in \F G\,|\, \psi(\alpha)=\alpha,\,\mbox{ for all }\,\psi \in {\mathcal{L}}\Aut(G)\}
$$
 and  the set of  primitive idempotents of $\mathcal{A}$. 
\end{lemma}
\begin{IEEEproof}
For an element $g\in G$, set $$\gamma_g=\displaystyle\sum_{\stackrel{\gcd ( i, o(g))=1}{0\leq i\leq o(g)}} g^i.$$
We claim that $\gamma_g\in\mathcal{A}$, for all $g\in G$.
Indeed, by Lemma~\ref{locautchar}, for each $\psi\in {\mathcal{L}}\Aut(G)$, there exists $r\in\mathbb{N}$ such that
$\gcd ( r,\,|G|) =1$ and $\psi(h)=h^{r}$, for all $h\in G$. Hence $$\psi(\gamma_g)=\displaystyle
\sum_{\stackrel{\gcd ( i, o(g))=1}{0\leq i\leq o(g)}} g^{ir}=\gamma_g.$$ 

It is easy to see that given two elements $g_1,\,g_2\in G$, either $\gamma_{g_1}=\gamma_{g_2}$ or they have disjoint supports in $\F G$.

Let $\Gamma=\{ g_1, g_2,\ldots, g_s\}$ be a complete set of elements in $G$ such that $\gamma_{g_i}\neq \gamma_{g_j}$, for $i\neq j$.

We claim that
$\{ \gamma_g\,|\, g\in \Gamma \}$ is an $\F$-basis of $\mathcal{A}$.
Indeed, by considering the respective supports, it is clear that
the set $\{ \gamma_g\,|\, g\in \Gamma\}$ is linearly independent.
Now given $\alpha\in{\mathcal{A}}$ and $g\in G$,
the coefficients of $g$ and $g^r$ in $\alpha$, for all $r\in \mathbb{N}$ with $\gcd ( r,\,o(g))=1$,
must all be equal, by Lemma~\ref{gemgr}. Thus, we may write:
$$
\alpha = a_{g_1}\gamma_{g_1} + a_{g_2}\gamma_{g_2} + \cdots + a_{g_s}\gamma_{g_s},
$$
hence $\{ \gamma_g\,|\, g\in \Gamma\}$ is a basis for $\mathcal{A}$.

Notice that there exists a bijection between the set of cyclic subgroups of $G$ and the set $\{ \gamma_g\,|\, g\in \Gamma\}$.
By Theorem~\ref{annihil}, 
there exists also a bijection between the set of the cyclic subgroups of $G$ and ${\mathcal{S}}_{cc}(G)$, hence
$\dim\mathcal{A}=|{\mathcal{S}}_{cc}(G)|$.

By Lemma~\ref{invaut}, it is clear that ${\mathcal{B}}\subset \mathcal{A}$.
In order to prove that ${\mathcal{B}}$ is linearly independent in $\F G$, assume that 
\begin{equation}\label{sumak}
\sum_{K\in {\mathcal{S}}_{cc}(G)} a_K\, e_K = 0,
\end{equation}
with $a_K\in\F$. Multiplying~\eqref{sumak} by $e_H$, for a fixed $H \in {\mathcal{S}}_{cc}(G)$, we get $a_H\,e_H = 0$, implying $a_H=0$, for all
$H \in {\mathcal{S}}_{cc}(G)$. Therefore, $\mathcal{B}$ is linearly independent in $\F G$ and thus a basis for $\mathcal{A}$.

For an element $e_H\in {\mathcal{B}}$, assume, by way of contradiction, that 
$e_H=e_1+e_2$, with $e_1$ and $e_2$ orthogonal idempotents of $\mathcal{A}$.
As ${\mathcal{B}}$ is a basis for ${\mathcal{A}}$, we can write
$$e_{i}=\sum_{j=1}^{n_i}b_{ji}e_{H_j},\;\mbox{ with } e_{H_j}\in\mathcal{B},\; b_{ij}\in\F, 1\leq j\leq n,\; i=1,2. $$

By Proposition~\ref{baseB} we have either $e_{H_{j_0}}e_H=e_H$ (when $H_{j_0}=H$) or $e_{H_{j}}e_H=0$, for $j\neq j_0$. Hence
$e_ie_H=b_{j_0}e_H$, if $H_{j_0}=H$, for some $j_0$ or $e_ie_H=0$, otherwise, $i=1,2$. As $e_H=e_H(e_1+e_2)$, we have that 
either $e_1=0$ or $e_2=e_H$ or $e_1=e_H$ and $e_2=0$. Therefore, $e_H$ is primitive in $\mathcal{A}$.

To finish the proof, write $1=\displaystyle\sum_{H\in {\mathcal{S}}_{cc}(G)} a_H e_H$, with $a_H\in \F$. 
For $K\in {\mathcal{S}}_{cc}(G)$,
multiply by $e_K$ and get $e_K=\displaystyle\sum_{H\in {\mathcal{S}}_{cc}(G)} a_H e_Ke_H=a_Ke_K$ implying
$a_K=1$, for all $K\in {\mathcal{S}}_{cc}(G)$. 

\end{IEEEproof}

\begin{remark}
Notice that Lemma~\ref{basepA} shows that $$\mathcal{A}=\displaystyle\bigoplus_{e_H\in\mathcal{B}} \F \cdot e_H.$$
\end{remark}

\begin{prop}\label{prop2}
Let $G$ be a finite abelian group and $\F$ a field such that $\car(\F)\not |\;|G|$.
If $e_1,\,e_2\in \mathcal{P}(\F G)$ and $H_{e_1}=H_{e_2}$, then there exists an automorphism
$\psi\in \mathcal{L}\Aut(G)$ whose linear extension to $\F G$ maps $e_1$ to $e_2$.
\end{prop}
\begin{IEEEproof}
Let $H\in {\mathcal{S}}_{cc}(G)$ be such that $e_1e_H=e_1$, that is, $H_{e_1}=H$. By Theorem~\ref{somaeH} have
\begin{equation}\label{eHsoma}
e_H=\sum_{H_{f}=H} f,\quad\mbox{ where } f\in \mathcal{P}(\F G).
\end{equation}
By Proposition~\ref{prop1}, for any $\psi\in {\mathcal{L}}\Aut(G)$, we have
\begin{equation}\label{e1H}
\psi(e_1) \psi(e_H)=\psi(e_1) e_{\psi(H)}  =\psi(e_1) e_H = \psi(e_1).
\end{equation}

Let
\begin{equation}\label{e0}
e_0=\sum \psi(e_1),
\end{equation}
where the sum runs over all distinct images of $e_1$ by $\psi\in\mathcal{L}\Aut(G)$. Then
$e_0$ is an idempotent and $\varphi(e_0)=e_0$, for all $\varphi \in {\mathcal{L}}\Aut(G)$, that is, $e_0\in\mathcal{A}$.
Hence, by Lemma~\ref{basepA},
$$
e_0=\sum_{e_H\in\mathcal{B}} a_H\,e_H.
$$
By construction of $e_0$ and~\eqref{e1H}, we have $e_0e_H=e_0$ and
$e_0e_K=0$, for all $K\in{\mathcal{S}}_{cc}(G)$ with $K\neq H$. As $e_H$ is a primitive idempotent of $\mathcal{A}$, it follows
 $e_0=e_H$.

Now comparing the expression of $e_H$ in~\eqref{eHsoma} and~\eqref{e0}, by the uniqueness of the sum of primitive idempotents, each $f$ in the sum~\eqref{eHsoma} is such that $f= \psi(e_1)$, for some $\psi\in  {\mathcal{L}}\Aut(G)$. As $e_2$ is one of the idempotents 
in~\eqref{eHsoma}, the result follows.
\end{IEEEproof}

\begin{prop}\label{recprop1}
Let $G$ be a finite abelian group and $\F$ a field such that $\car(\F)\not\! |\;|G|$.
If $e_1,\,e_2\in \mathcal{P}(\F G)$ are such that $\psi(H_{e_1})=H_{e_2}$, for some $\psi\in \Aut(G)$, then
there exists an automorphism $\theta\in \Aut(G)$ whose linear extension to $\F G$ maps  $e_1$ and $e_2$, i.e., the ideals 
of $\F G$  generated by
$e_1$ and $e_2$ are $G$-equivalent.
\end{prop}
\begin{IEEEproof}
Since $\psi(H_{e_1})=H_{e_2}$, for $\psi\in \Aut(G)$, by Lemma~\ref{invaut}, we have
$$
\psi(e_1)\,e_{H_{e_2}}\,=\,\psi(e_1)\,\psi(e_{H_{e_1}})\,=\,\psi(e\,e_{H_{e_1}})\,=\,\psi(e_1).
$$
Hence, by the uniqueness, we have $H_{\psi(e_1)}=H_{e_2}$. Now, by Proposition~\ref{prop2}, there exists a local automorphism
$\delta\in {\mathcal{L}}\Aut(G)$ such that $\delta(\psi(e_1))=e_2$. Therefore, taking $\theta=\delta\psi\in \Aut(G)$, the result follows.
\end{IEEEproof}

\section{Applications to Coding Theory}\label{positive}

We denote by $C_n$ the cyclic group of order $n$.
In the context of Coding Theory, the following results appear in~\cite{miller}.

\

\noindent{\bf Theorem A}~\cite[Theorem 3.6]{miller}  {\it Let  $G$ be a finite abelian group of odd 
order and exponent $n$ and denote by
$\tau(n)$ the number of divisors of $n$. Then there exist  precisely $\tau(n)$ non $G$-equivalent minimal abelian codes 
 in $\F_2 G$}.

\

\noindent{\bf Theorem B}~\cite[Theorem 3.9]{miller}  {\it Let  $G$ be a finite abelian group of odd order. Then two minimal abelian codes in $\F_2 G$
 are $G$-equivalent if and only if they have the same weight distribution.}

\

Unfortunately these statements are not correct. The errors arise from the assumption, implicit in the last paragraph of~\cite[p. 167]{miller}, that
if  $e$ and $f$ are primitive idempotents of $\F_2 C_m$ and $\F_2 C_n$, respectively, then $ef$ is a primitive idempotent of
$\F_2[C_m\times C_n]$. 

We exhibit below counterexamples to both Theorems A and B that were first communicated in~\cite{FGM}.
However,  Theorem A does hold under certain restrictive hypotheses, as we show in the next section.

\begin{prop} \label{cp2cp}
Let $p$ be an odd prime such that $\bar 2$ generates $U(\zzz_{p^2})$ and $G=\left< a\right>\times \left< b\right>$ an abelian group, with  \linebreak $o(a)=p^2$ and $o(b)=p$. Then $\F_2G$ has four inequivalent minimal codes, namely, the ones generated by the idempotents
$e_0=\wh G$, $e_1= \widehat{b} - \widehat{\left< a^p\right>\times \left< b\right>} $,
$e_2= \widehat{a} - \wh G$ and $e_3= \widehat{\left< a^p\right>\times \left< b\right>} - \wh G.$

Also all minimal codes  of $\F_2 G$ are described in the following table with their dimension and weight.

\

$$
{\small
\begin{array}{|l|c|c|c|} \hline \label{cp2cptable}
\mbox{\rm Code} & \mbox{\small\rm Primitive Idempotent} & \mbox{\rm Dimension} & \mbox{\small\rm Weight}  \\
             \hline\hline
I_0 & e_0=\wh{a}\wh{b} = \wh G       &  1      & p^3  \\ \hline
I_{1} & e_1= \wh{b} - \wh{\left< a^p\right>\times \left< b\right>}       &  p^2-p      & 2p   \\ \hline
I_{1j} & e_{1j}= \wh{a^{jp}b} - \wh{\left< a^p\right>\times \left< b\right>}       &  p^2-p      & 2p  \\
&   j=1,\ldots, p-1 & & \\ \hline
I_{2} & e_2=\wh{a} - \wh{G}       &  p-1      & 2p^2   \\ \hline
I_{2i} & e_{2i}=\wh{ab^i} - \wh{G}       &  p-1      & 2p^2  \\
& i=1,\ldots, p-1 & & \\ \hline
I_{3} & e_3=\wh{\left< a^p\right>\times \left< b\right>} - \wh G        & p-1      & 2p^2  \\ \hline
    \hline
\end{array}
 } 
$$
\end{prop}

\

\begin{IEEEproof}
In order to use Theorem~\ref{3.4}, first we need to find all subgroups $H$ of $G$ such that $G/H$ is cyclic.
Notice that the $p+1$ distinct subgroups of order $p^2$ of $G$ are $\left< ab^i\right>$, for $i=0,\ldots , p-1$, 
and $\left< a^p\right>\times \left<b \right>$. The $p+1$ distinct subgroups of
order $p$ of $G$ are $\left< a^{jp}b\right>$, for $j=0,\ldots , p-1$, and $\left< a^p\right>$. 
The subgroups $\left< a^{jp}b\right>$, for all $j=0,\ldots , p-1$, are contained only in $\left< a^p\right>\times 
\left< b\right>$ and $\left< a^p\right>$ is contained in all
subgroups of order $p^2$.  Besides, all quotients of $G$ by these subgroups are cyclic, except $G/\left< a^p\right>$ 
which is the unique noncyclic quotient of $G$. The quotient of $G$ by $\left< a^p\right>\times \left< b\right>$ is also cyclic.
\vspace{0.2cm}

Now applying Theorem~\ref{3.4}, we have the following minimal codes generated by primitive idempotents.

The code $I_0 = \F_2G\cdot e_0$, where $e_0=\wh G$ and $\dim I_0 = 1$.

As $\left< b\right>$ is uniquely contained in $\left< a^p\right>\times \left< b\right>$, we have $I_1 =  \F_2G\cdot e_1$, where $e_1= \widehat{b}-\widehat{\left< a^p\right>\times \left< b\right>} $, and
$\dim I_1 = \phi(p^2)=p^2-p$.
The codes $I_{1j} = \F_2G\cdot e_{1j}$, where $e_{1j}=\widehat{a^{jp}b} - \widehat{\left< a^p\right>\times \left< b\right>} $, for
all $j=1,\ldots, p-1$, are all equivalent to $I_1$, since the extension to the group algebra $\F_2G$
of the isomorphism  $\psi_j : G\rightarrow G$ given by
$\psi(a)=a$ and $\psi(b)=a^{jp}b$, for each $j$, maps $I_{1}$ onto $I_{1j}$.

Let $I_2=\F_2G\cdot e_2$, where $e_2=  \widehat{a} - \wh G$, and
$I_3=\F_2G\cdot e_3$, where $e_3=\widehat{\left< a^p\right>\times \left< b\right>} - \wh G$. We have $\dim I_3=\dim I_2 = \phi(p)=p-1$.
We also have the codes $I_{2i}=\F_2 G\cdot e_{2i}$, where $e_{2i}=\widehat{ab^i} -\wh G $, for $i=1,\ldots, p-1$, all equivalent to
$I_2$ with  corresponding isomorphism $\varphi_i : G\rightarrow G$ given by $\varphi(a)=ab^i$ and $\varphi(b)=b$.

\vspace{0.2cm}

We prove now
that the codes $I_k$, with $k=0,1,2,3$, are four inequivalent minimal codes in $\F_2G$.

It is obvious that $I_0$ is not equivalent to any of the other codes $I_k$, for $k\neq 0$, and also that
$I_1$ is not equivalent to either $I_2$ or $I_3$. Let us prove that $I_2$ and $I_3$ are inequivalent.

Notice that $supp(e_2)= G\setminus\left< a\right>$, which contains elements of order $p$, and $supp(e_3)=G\setminus\left< a^p\right>\times \left< b\right>$, which only contains elements of order $p^2$. Hence, if there is an isomorphism \linebreak $\psi : G\rightarrow G$ such that
 $\psi(e_2)=e_3$,
we would have elements of order $p$ being mapped to elements of order $p^2$, a contradiction. Therefore, $I_2$ is not equivalent to $I_3$.

\vspace{0.2cm}

It is clear that the minimal code $I_0$ has weight $p^3$, as all its nonzero elements have this weight.

For $1\leq j\neq k\leq p-1$, as $supp(a^{jp}\wh b)\cap supp(a^{kp}\wh b)=\emptyset$, the element $(a^{jp}+a^{kp})e_1=(a^{jp}+a^{kp})\wh b$ is in $I_1$ and has weight $2p$.
Notice that $I_1\subset \F_2G\cdot \wh b$, thus the weight of any element of $I_1$ must be a multiple of $p$.
Hence, if there is an element in $I_1$ of weight $p$, it should be of the form $a^i\wh b$. But
$a^i\wh b\cdot e_1 = a^i(a^p+a{2p}+\cdots +a^{(p-1)p})\wh b\neq a^i\wh b$ which implies $a^i\wh b\not\in I_1$, for any $1\leq i$.
Therefore, the  weight of $I_1$ is $2p$.

The  weights of $I_2$ and $I_3$ will follow from the proof of the next proposition.
\end{IEEEproof}

\begin{prop} \label{samedistrib}
The (inequivalent) minimal codes $I_2$ and $I_3$ of Proposition~\ref{cp2cp} have the same weight distribution.
\end{prop}
\begin{IEEEproof}
An $\F_2$-basis for the code $I_2$ is $$\beta = \{\mu_i = (\hat G - \widehat{a})b^i =  \hat G - \hat a b^i | 1\leq i\leq p-1\}.$$
For $1\leq i\neq j\leq p-1$, we have $supp(\hat a b^i) \cap supp(\hat a b^j) = \emptyset$. Hence, for an element $\alpha\in I_2$,
we have:

\

{\bf Case 1.} $\alpha$ is a sum of an even number of $\mu_i$'s. Thus
$$ \alpha   =  \mu_{i_1}+\cdots + \mu_{i_{2k}}  
= 2k\wh G - \hat a ( b^{i_1} +\cdots + b^{i_{2k}}),$$
which is an element of weight $2kp^2$. Besides, in $I_2$ we have at least $\left( \begin{array}{c} p-1 \\ 2k \end{array}\right)$ distinct elements with weight
$2kp^2$.

\

{\bf Case 2.} $\alpha$ is a sum of an odd number of $\mu_i$'s. Thus, for $k'\geq 1$,
$$ \alpha   =  \mu_{i_1}+\cdots + \mu_{i_{2k'-1}} 
= \hat G - \hat a ( b^{i_1} +\cdots + b^{i_{2k'-1}}),$$
which is an element of weight $$p^3-(2k'-1)p^2 = p^2(p-2k'+1)= 2kp^2$$ (where $k'=\frac{p+1-2k}{2}$). Hence, in $I_2$, there are $\left( \begin{array}{c} p-1 \\ 2k'-1 \end{array}\right) = \left(\begin{array}{c} p-1 \\ p-2k \end{array}\right)$ distinct elements with weight
$2kp^2$.

Therefore, for each $k\geq 1$, there are $\left( \begin{array}{c} p-1 \\ 2k \end{array}\right) + \left( \begin{array}{c} p-1 \\ p-2k \end{array}\right) = \left( \begin{array}{c} p \\ 2k \end{array}\right)$ elements of weight $2kp^2$ in $I_2$.

\

Similarly, an $\F_2$-basis for the code $I_3$ is $$\gamma = \{\delta_i = e_3a^i=\widehat G - \widehat{a^p}\widehat{b}a^i | 1\leq i\leq p-1\}.$$
For $1\leq i\neq j\leq p-1$, $supp(\wh{a^p}\widehat{b}a^i ) \cap supp(\wh{a^p}\widehat{b}a^j) = \emptyset$. Hence, for an element $\alpha\in I_3$,
we have:

\

{\bf Case 1.} $\alpha$ is a sum of an even number of $\delta_i$'s. Thus
$$ \alpha   =  \delta_{i_1}+\cdots + \delta_{i_{2k}}
= 2k\hat G - \widehat{a^p}\widehat{b} ( a^{i_1} +\cdots + a^{i_{2k}}),$$
which is an element of weight $2kp^2$. Besides, in $I_3$ we have at least $\left( \begin{array}{c} p-1 \\ 2k \end{array}\right)$ distinct such
elements with weight $2kp^2$.

\

{\bf Case 2.} $\alpha$ is a sum of an odd number of $\delta_i$'s. Thus, for $k'\geq 1$,
$$ \alpha   =  \delta{i_1}+\cdots + \delta_{i_{2k'-1}}
= \wh G - \widehat{a^p}\widehat{b} ( a^{i_1} +\cdots + a^{i_{2k'-1}}),$$
which is an element of weight $$p^3-(2k'-1)p^2 = p^2(p-2k'+1)= 2kp^2,$$ where $k'=\frac{p+1-2k}{2}$. Hence, in $I_3$, there are  $\left( \begin{array}{c} p-1 \\ 2k'-1 \end{array}\right) = \left(\begin{array}{c} p-1 \\ p-2k \end{array}\right)$ distinct elements with weight
$2kp^2$.

Therefore, also in $I_3$, for each $k\geq 1$, there are $\left( \begin{array}{c} p-1 \\ 2k \end{array}\right) + \left( \begin{array}{c} p-1 \\ p-2k \end{array}\right) = \left( \begin{array}{c} p \\ 2k \end{array}\right)$ elements of weight $2kp^2$.

As $\displaystyle\sum_{k=1}^{(p-1)/2}\left( \begin{array}{c} p \\ 2k \end{array}\right) = 2^{p-1}$,
this proves that the weight distribution of $I_2$ and $I_3$ are the same, but $I_2$ and $I_3$ are not equivalent. Besides,
the weight of these codes is $2p^2$.
\end{IEEEproof}

\

Observe that the group $G$ of Proposition~\ref{cp2cp} has exponent $p^2$ and $\tau(p^2)=3$, however, $\F_2 G$ has four inequivalent minimal
 codes. This is a counterexample to Theorem~A.

\vspace{0.2cm}

Notice that Proposition~\ref{samedistrib} actually exhibits a counterexample also to Theorem~B.

\vspace{0.2cm}

In the following proposition, we study the minimal codes in $\F_2(C_{p^n}\times C_{p})$, for an odd prime $p$ and $n\geq 3$.
Its proof is similar to the proof of Proposition~\ref{cp2cp}.
This gives a whole family of counterexamples to Theorem~A.

\

\begin{prop}\label{codescpncp}
Let $n\geq 3$ be a positive integer and $p$ an odd prime such that $\bar 2$ generates
$U(\zzz_{p^2})$ and $G=\left< a\right>\times \left< b\right>$ be an abelian group, with $o(a)=p^n$ and $o(b)=p$.
Then the minimal codes of $\F_2 G$ are described in the following table.
$$
{\small
\begin{array}{|l|c|c|} \hline \label{cpncptable}
\mbox{\rm Code} & \mbox{\rm Dimension} & \mbox{\rm Weight}  \\
       \hline\hline
I_0 =\langle\wh{a}\wh{b}\rangle = \langle\wh G\rangle       &  1      & p^{n+1} \\ \hline
I_{1} =\langle\wh{\left< a^p\right>\times \left< b\right>} - \wh{G} \rangle     &  p-1      & 2p^n   \\ \hline
I_{1i} =\langle \wh{ab^i} - \wh{G}\rangle & p-1 & 2p^n \\
 i=0,\ldots, p-1 & &  \\ \hline
I_{2} =\langle\wh{\left< a^{p^2}\right>\times \left< b\right>} -\wh{\left< a^p\right>\times \left< b\right>}\rangle &  p(p-1) & 2p^{n-1}   \\ \hline
I_{2i} =\langle\wh{a^{p}b^i} - \wh{\left< a^p\right>\times \left< b\right>}    \rangle &  p(p-1)      & 2p^{n-1} \\
 i=1,\ldots, p-1 &  & \\ \hline
\ldots     & \ldots & \\ \hline
I_{k} =\langle\wh{\left< a^{p^k}\right>\times \left< b\right>} -   \wh{\left< a^{p^{k-1}}\right>\times \left< b\right>}\rangle & p^{k-1}(p-1) & 2p^{n-k+1}  \\ \hline
I_{ki} =\langle\wh{a^{p^{k-1}}b^i} - \wh{\left< a^{p^{k-1}}\right>\times \left< b\right>}\rangle     &  p^{k-1}(p-1)      & 2p^{n-k+1}  \\
i=1,\ldots, p-1 & & \\ \hline
\ldots &     \ldots  & \\ \hline
I_{n-1} =\langle\wh{\left< b\right>} -   \wh{\left< a^{p^{n-2}}\right>\times \left< b\right>}\rangle & p^{(n-1)}(p-1) & 2p  \\ \hline
I_{n-1,i} =\langle\wh{a^{p^{n-1}}b^i} - \wh{\left< a^{p^{n-2}}\right>\times \left< b\right>}  \rangle   &  p^{(n-1)}(p-1)      & 2p \\
  i=1,\ldots, p-1 & & \\ \hline
\hline
\end{array}
} 
$$
There are $2n$ inequivalent minimal codes in $\F_2(C_{p^n}\times C_p)$.
\end{prop}

The following table presents the correspondence between the classes of $G$-isomorphisms of subgroups and
the $G$-equivalence classes of minimal codes in $\F_2(C_{p^n}\times C_{p})$, listing representatives of these classes.
$$
{\small
\begin{array}{|c|c|} \hline \label{cpncpgrp}
\mbox{\rm Subgroups } & \mbox{\rm Codes}  \\ \hline\hline
G & I_0 = \langle \wh G \rangle \\ \hline
\left< a \right> & I_{11}= \langle \wh a  - \wh{G} \rangle \\ \hline
\left< a^p \right> \times \left< b \right> & I_{1}= \langle \wh{\left< a^p \right> \times \left< b \right>} - \wh{G} \rangle \\ \hline
\left< a^pb \right> &  I_{21} =\langle \wh{a^pb}-\wh{\left< a^p\right>\times \left< b\right>}  \rangle \\ \hline
\left< a^{p^2} \right> \times \left< b \right> &  I_{2} =\langle \wh{\left< a^{p^2} \right> \times \left< b \right>}-
\wh{\left< a^p\right>\times \left< b\right>}  \rangle \\ \hline
\ldots & \ldots \\ \hline
\left< a^{p^k}b \right> &  I_{k+1,1} =\langle \wh{a^{p^k}b}-\wh{\left< a^{p^k}\right>\times \left< b\right>}  \rangle \\ \hline
\left< a^{p^{k+1}} \right> \times \left< b \right> &  I_{k+1} =\langle \wh{\left< a^{p{k+1}} \right> \times \left< b \right>}-
\wh{\left< a^{p^k}\right>\times \left< b\right>}  \rangle \\ \hline
\ldots & \ldots \\ \hline
\left< b \right> &  I_{n-1} =\langle \wh b-
\wh{\left< a^{p^n{-1}}\right>\times \left< b\right>}  \rangle \\ \hline
   \hline
\end{array}
} 
$$

\section{A positive result for codes}
 
In~\cite{FGM} we showed that Theorem~A holds in the special case of minimal codes in $\F_2(C_{p^n}\times C_{p^n})$. 
We now prove this Theorem in a more general situation.

\begin{lemma}\label{maxcyclic}
If $H$ is a cyclic subgroup of order $p^s$ in a group \linebreak $G\cong \underbrace{C_{p^r}\times \cdots \times C_{p^r}}_{m}$, with $s\leq r$, then
there exists a cyclic subgroup of $G$, of order $p^r$, containing $H$. 
\end{lemma}
\begin{IEEEproof}
Write $G=\langle g_1\rangle \times \cdots \times \langle g_m\rangle$, with $o(g_i)=p^r$,  $1\leq i\leq m$.
 Since $H$ is a cyclic subgroup of $G$, we have $H=\langle g_1^{j_1p^{t_1}} g_2^{j_2p^{t_2}}\cdots g_m^{j_mp^{t_m}}\rangle $, with
 $\gcd (j_i, p)=1$  and $0\leq t_i\leq r$,  $1\leq i\leq m$.
 Then one of the exponents $t_k$,  $1\leq k\leq m$, is minimal and, for such $t_k$, we consider the element
 $a=g_1^{j_1p^{t_1-t_k}}\cdots g_k^{j_kp^{0}}\cdots g_m^{j_mp^{t_m-t_k}}\in G$. As   $\gcd (j_k, p)=1$, we have $o(a)=p^r$ and
 $\langle a^{p^{t_k}}\rangle = H$, as $G\cong \underbrace{C_{p^r}\times \cdots \times C_{p^r}}_{m}$.
\end{IEEEproof}

\begin{theo}\label{minHtcprxcpr}
Let $m$ and $r$ be positive integers. If $G=\underbrace{C_{p^r}\times \cdots \times C_{p^r}}_{m}$  is a finite abelian $p$-group, then
any co-cyclic subgroup of $G$ contains a subgroup isomorphic to
$\underbrace{C_{p^r}\times \cdots \times C_{p^r}}_{(m-1)}$. Hence the subgroups of $G$ isomorphic to
$\underbrace{C_{p^r}\times \cdots \times C_{p^r}}_{(m-1)}$ are precisely the minimal co-cyclic subgroups of $G$.
\end{theo}
\begin{IEEEproof}
Let $H\in {\mathcal{S}}_{cc}(G)$. Then, by Theorem~\ref{annihil}(3), $H^{\perp}$ is a cyclic subgroup of $G^*$.
By Theorem~\ref{annihil}(1) and Lemma~\ref{maxcyclic}, there exists $K^{\perp}\cong C_{p^r}$ a subgroup of $G^*$ such that $H^{\perp}\subset K^{\perp}$.
Then $K=\Psi^{-1}(K^{\perp})$ is a subgroup of $G$ such that $K\cong \underbrace{C_{p^r}\times \cdots \times C_{p^r}}_{(m-1)}$ and $K\subset H$, proving the theorem.
\end{IEEEproof}

As a consequence of the results above, we get:

\begin{prop}\label{propcprgeral}
Let $m$ and $r$ be positive integers. If $G=\underbrace{C_{p^r}\times \cdots \times C_{p^r}}_{m}$  is a finite abelian $p$-group
and $\F$ is a field with $ \car (\F)\neq p$, then a primitive idempotent of $\F G$ is of the form
$\wh K\cdot e_h$, where $K$ is a subgroup of $G$ isomorphic to
$\underbrace{C_{p^r}\times \cdots \times C_{p^r}}_{(m-1)}$ and $e_h$ is a primitive
idempotent of $\F\langle h\rangle$, where $h\in G$ is such that $G=\langle h\rangle \times K$ and $\langle h\rangle \cong C_{p^r}$.
\end{prop}
\begin{IEEEproof}
Let $e$ be a primitive idempotent of $\F G$ and $H$ be the unique co-cyclic subgroup of $G$ such that $e=\wh H\cdot e$,
by Theorem~\ref{unicoHe}. By Theorem~\ref{minHtcprxcpr}, there exists $K\cong \underbrace{C_{p^r}\times \cdots \times C_{p^r}}_{(m-1)}$
 contained in $H$ such that $e\wh K=e$.

Let $h\in G$ be such that $G/K=\langle Kh\rangle$.
Since $|G/K|=p^r$ and $exp(G)=p^r$, it follows directly that $o(h)=p^r$. Hence, we can write:
\begin{eqnarray}
e & = & \wh K\cdot e \nonumber \\
& = & \wh K \left[ \left( \sum_{g\in K}\alpha_{0,g} g \right)1 
+ \cdots + \left( \sum_{g\in K}\alpha_{p^{r}-1,g} g \right)h^{p^r-1}\right] \nonumber\\
& = & \wh K \left(\beta_0\cdot 1 + \beta_1\cdot h 
+ \cdots + \beta_{p^r-1}\cdot h^{p^r-1}\right),
\end{eqnarray}
where $\beta_i=\displaystyle\sum_{g\in K}\alpha_{i,g} \in \F$ and $h^t\not\in K$, $1\leq t\leq p^r-1$.

Let $\psi : G\longrightarrow G/K$ be the natural homomorphism.
Since $e = \wh K \cdot e$ is an idempotent in $\F G$, we have
\begin{eqnarray}
\psi(e) & = & \psi(\wh K\cdot e) \nonumber \\
& = & \psi \left( \wh K (\beta_0\cdot 1 + \beta_1\cdot h 
+ \cdots + \beta_{p^r-1}\cdot h^{p^r-1})\right) \nonumber \\
& = & \beta_0\cdot 1 + \beta_1\cdot h 
 + \cdots + \beta_{p^r-1}\cdot h^{p^r-1} \nonumber
\end{eqnarray}
is also an idempotent.

{\bf Claim:} $e_h=\psi(e) = \beta_0\cdot 1 + \beta_1\cdot h 
+ \cdots + \beta_{p^r-1}\cdot h^{p^r-1}$ is a primitive idempotent in $\F \langle h\rangle\cong \F (G/K)$.

Indeed, if $e_h=e_1+e_2$, with $e_1$ and $e_2$ orthogonal idempotents in $\F \langle h\rangle$, then
$$
e=\wh K\cdot e_h = \wh K\cdot e_1 + \wh K\cdot e_2,
$$
with $(\wh K\cdot e_i)^2=\wh K\cdot e_i$, for $i=1,2$, and $\wh K\cdot e_1\wh K\cdot e_2 = 0$, as $e_1\cdot e_2 =0$.
Since $e$ is a primitive idempotent, then either $\wh K\cdot e_1=0$ or $\wh K\cdot e_2=0$.
Suppose $\wh K\cdot e_1=0$ and write $ e_1= \gamma_0\cdot 1 + \gamma_1\cdot h + \gamma_2 \cdot h^2 + \cdots + \gamma_{p^r-1}\cdot h^{p^r-1}$.
Then
\begin{eqnarray*}\label{gamas}
0 & = &
\wh K\cdot e_1 \\ 
& = & \frac{1}{|K|}\left(\sum_{k\in K} k\right) (\gamma_0 + \gamma_1\cdot h 
+ \cdots + \gamma_{p^r-1}\cdot h^{p^r-1}) \\
& = & \frac{1}{|K|} \left[\left(\sum_{k\in K}\gamma_0 k\right)
+ \cdots + \left(\sum_{k\in K}\gamma_{p^r-1} k\right)h^{p^r-1}\right]
\end{eqnarray*}
implies $\gamma_i=0$,  $0\leq i\leq p^r-1$, as the summands in~\eqref{gamas} have disjoint supports in $\F G$. This proves the proposition.
\end{IEEEproof}

These results can be applied and extended as follows.

\begin{cor}
Let $m$ and $r$ be positive integers, a finite abelian $p$-group 
$G= \underbrace{C_{p^r}\times \cdots \times C_{p^r}}_{m}$  and $\F_q$
a finite field with $q$ elements such that \linebreak $o(\bar{q})=\phi(p^r)$ in $U(\zzz_{p^r})$. Then
 the minimal abelian codes  in $\F_q G$ are as follows:
$$
\begin{array}{|c|c|c|}
\hline\hline
\mbox{\rm Primitive Idempotent} &  \mbox{\rm Dimension} & \mbox{\rm Weight}\\ \hline\hline
\wh G     & 1  & p^{rm}\\ \hline
 \wh{K}(\wh{h^{p}} - \wh{h}) & p-1 &  2p^{r(m-1)+(r-1)}\\ \hline
 \wh{K}(\wh{h^{p^2}} - \wh{h^{p}}) & p(p-1) &  2p^{r(m-1)+(r-2)}\\ \hline
 \wh{K} (\wh{h^{p^3}} - \wh{h^{p^2}}) & p^2(p-1) & 2p^{r(m-1)-(r-3)} \\ \hline
\cdots  & \cdots &  \\ \hline
\wh{K}(\wh{h^{p^{i}}} - \wh{h^{p^{i-1}}}) & p^{i-1}(p-1) & 2p^{r(m-1)-(r-i)}\\ \hline
\cdots  & \cdots & \\ \hline
 \wh{K} (1 - \wh{h^{p^{r-1}}}) & p^{r-1}(p-1) & 2p^{r(m-1)} \\
 \hline\hline
\end{array}
$$
where $h$ is as in Proposition~\ref{propcprgeral}.
Consequently, the number of non $G$-equivalent minimal abelian codes is $r+1=\tau(p^r)$.
\end{cor}

\begin{cor}
Let $n\geq 2$ be an integer, $G=\underbrace{C_n\times \cdots \times C_n}_{m}$ an abelian group and $\F_q$ a finite field
such that $\gcd ( q, n) = 1$.
Then the primitive idempotents of $\F_qG$ are of the form $\wh K\cdot e_h$, where $K$ is a subgroup of $G$ isomorphic to
$\underbrace{C_n\times \cdots \times C_n}_{(m-1)}$, $h\in G$ is such that $G=K\times \langle h\rangle$ and $e_h$ is a primitive
idempotent of $\F_q\langle h\rangle$.
\end{cor}
\begin{IEEEproof}
Let $n=p_1^{n_1}p_2^{n_2}\cdots p_t^{n_t}$, with $p_i$ rational primes, $1\leq i\leq t$, 
and $p_i\neq p_j$, for $i\neq j$ and $n_i\geq 1$. Let $G_i=\underbrace{C_{p_i^{r_i}}\times\cdots \times C_{p_i^{r_i}}}_{m}$ be the $p_i$-Sylow
subgroup of $G$.

Let $0\neq e\in \F_qG$ be a primitive idempotent. By Lemma~\ref{uniqueco-cyc}, there exists a unique $H\in  {\mathcal{S}}_{cc}(G)$
such that $e\cdot e_H = e$ and $e\cdot e_K=0$, for any other $K\in {\mathcal{S}}_{cc}(G)$, with $K\neq H$.
By~\eqref{idempH}, $e_H = e_{H_{1}}e_{H_{2}}\cdots e_{H_{t}}$, where $e_{H_{i}}=\wh{G_{i}}$ or $e_{H_{i}}=\wh{H_{i}}-\wh{H_{i}^{\sharp}}$,
for $H_{i}$ the $p_i$-Sylow of $H$. Hence, as $0\neq e$ is primitive, we have
$$
e\cdot e_H = e\qquad \Rightarrow \qquad e\cdot e_{H_i} = e.
$$
By Proposition~\ref{unicoHe} and Theorem~\ref{minHtcprxcpr}, there exists $K_i=\underbrace{C_{p_i^{r_i}}\times\cdots \times C_{p_i^{r_i}}}_{(m-1)}$
a subgroup of $G_i$ such that $K_i\leq H_i$ and
$e=e\cdot \wh{K_i}$, for all $1\leq i \leq t$. Hence
\begin{eqnarray*}
e & = & e\cdot \wh{K_1}=e\cdot \wh{K_2}\wh{K_1} \\
& & \cdots \\ 
& = & e\cdot\wh{K_t}\cdots\wh{K_2}\wh{K_1}= e\wh{K_t\cdots K_2K_1} \\ 
& \cong & e ({\underbrace{\wh{C_n\times \cdots \times C_n}}_{(m-1)\mbox{ factors}}}).
\end{eqnarray*}
Let $K=\underbrace{C_n\times \cdots \times C_n}_{(m-1)\mbox{ factors}}$. Then $G/K\cong C_n$. Take $a\in G$ such that $G/K=\langle Ka\rangle$.
Similarly to the proof of Theorem~\ref{propcprgeral}, we show that $o(a)=n$ and that $\psi(e)$ is a primitive idempotent of $\F_q\langle a\rangle$, where $\psi : \F_qG\longrightarrow \F_q(G/K)$.
\end{IEEEproof}

Using the results from this section and from Section~\ref{sec3}, we obtain the following.

\begin{theo}
Let $G$ be a finite abelian group of exponent $n$ and $\F$ a finite field such that $char(\F)\not |\, |G|$.
Then the number of non $G$-equivalent minimal abelian codes is precisely $\tau(n)$ if and only if $G$ is a direct product
of cyclic groups isomorphic to one another.
\end{theo}

\ifCLASSOPTIONcaptionsoff
  \newpage
\fi

\end{document}